\documentclass[]{article}

\usepackage{amsmath}
\usepackage{amssymb}

\usepackage{graphicx}

\usepackage{cite}
\usepackage{color} 

\usepackage{subfigure}
\usepackage{multicol}
\usepackage{array}

\begin{document}
%\begin{frontmatter}

\title{A population-based microbial oscillator}
\date{}

% Insert Author names, affiliations and corresponding author email.
\author{
Angel Go\~ni-Moreno$^{1,\ast}$ and Martyn Amos$^{2}$ \\ \vspace{8pt} \\
\small $^{1}$ Natural Computing Group, Universidad Polite\'cnica de Madrid, Spain \\
\small $^{2}$ Department of Computing and Mathematics, \\ \small Manchester Metropolitan University, UK \\
\small $\ast$ E-mail: agmoreno@gcn.upm.es, m.amos@mmu.ac.uk
}

\maketitle

\begin{abstract}
Genetic oscillators are a major theme of interest in the emerging field of synthetic biology. Until recently, most work has been carried out using intra-cellular oscillators, but this approach restricts the broader applicability of such systems. Motivated by a desire to develop large-scale, spatially-distributed cell-based computational systems, we present an initial design for a {\it population-level} oscillator which uses three different bacterial strains. Our system is based on the client-server model familiar to computer science, and uses quorum sensing for communication between nodes. We present the results of extensive {\it in silico} simulation tests, which confirm that our design is both feasible and robust.
\end{abstract}
%\end{frontmatter}

\section{Introduction}

The growing field of synthetic biology \cite{Benner:NatRevGenet:2005,Purnick:NatRevMolCellBiol:2009,serrano2007synthetic} has the potential to impact on many pressing areas of concern, such as health \cite{Lu:ProcNatlAcadSciUSA:2009,Ro:Nature:2006}, energy \cite{Lee:CurrOpinBiotechnol:2008} and the environment \cite{Sayler:CurrOpinMicrobiol:2004}. By engineering bacteria (and sometimes other types of cell), practitioners in the field hope to take advantage of their inherent ``biological nanotechnology". This engineering is generally achieved by modifying the natural transcriptional mechanisms and regulatory activities of the bacterium of interest. Collections of bacterial cells have recently been successfully engineered to perform simple tasks, such as emulating light-sensitive film \cite{Levskaya:Nature:2005}, or generating simple patterns \cite{Basu:Nature:2005,Sohka:JBiolEng:2009}. By harnessing and controlling communication and synchronization mechanisms found in such systems, we hope to engineer scalable, robust, fault-tolerant bacterial devices.

Our objective is to design a multi-strain bacterial community with autonomous behaviour. We model our system on the ``client-server" architecture familiar to computer science \cite{berson1996client}, with a single central server and two clients (one ``red" and the other ``green"). The task we define is that of {\it oscillation}; by engineering feedback between three different strains, web obtain indefinite switching between ``red" and ``green" outputs. 

In this paper we first briefly review previous work on engineered cellular oscillators. We then describe the architecture of the two clients and the single server strain, which all use standard genetic parts. We describe  {\it in silico} component testing results, before demonstrating, using extensive simulation studies, the feasibility of engineering multi-strain, {\it population-based oscillators}. Our results suggest that such distributed computations may become more common as the field of synthetic biology matures. 

\section{Previous work}

Synthetic biology is an emerging scientific discipline largely concerned with the engineering of biological systems. The goals of synthetic biology include the ``optimization” of existing biological systems for human purposes, and the development and application of rational engineering principles to the design and construction of biological systems. A significant amount of work on synthetic biology has concerned {\it switches} \cite{gardner2000construction} and {\it oscillators}; here, we focus on the latter. 

In physics, an oscillator is a system that produces a regular, periodic ``output". Familiar examples include a pendulum or a vibrating string. Linking several oscillators together in some way gives rise to {\it synchrony} -- for example, heart cells repeatedly firing in unison, or millions of fireflies blinking on and off, seemingly as one \cite{strogatz}. 

Although synthetic genetic oscillators date back to the early 1960s \cite{goodwin1963temporal}, these so-called {\it Goodwin oscillators} were limited to single genes. The archetype of the {\it multi-gene} oscillator is known as the {\it repressilator}, which is a ``ring" of genes, each repressing its successor \cite{Elowitz:Nature:2000,muller2006generalized}. A detailed discussion of synthetic genetic oscillators is beyond the scope of this paper, but we refer the reader to a recent extensive survey \cite{purcell2010comparative}.

Recently, genetic clocks have been coupled to produce synchronised oscillations at the level of a cell {\it population} \cite{Danino:Nature:2010}. Following on from earlier theoretical work \cite{GarciaOjalvo:ProcNatlAcadSciUSA:2004,mcmillen2002synchronizing}, this paper demonstrated the feasibility of engineering population-level oscillations. However, the population used was {\it homogenous}. In nature, there exist bacterial communities known as {\it biofilms} \cite{davies1998involvement}, in which hundreds of bacterial species form a robust and stable community through signalling and cooperation. If the potential of synthetic biology is to be fully realised, we believe that it is important to understand how to engineer communication in {\it mixed} groups of cells. We therefore describe a scheme for obtaining population-level oscillations using a {\it three-component} population.

\newpage
\section{A multi-strain bacterial oscillator}

\begin{figure}[]
\begin{centering}
\includegraphics[scale=0.35]{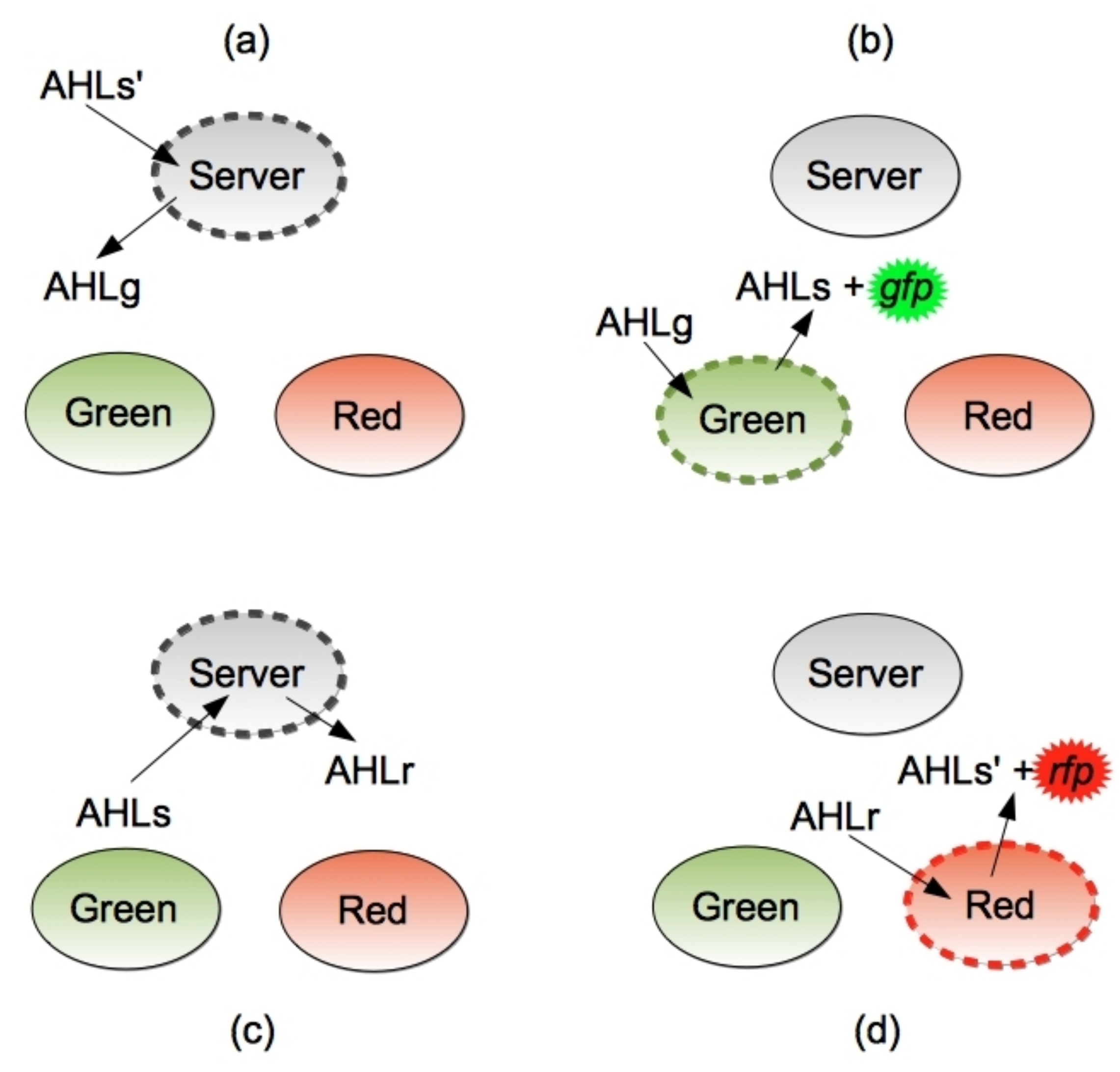}
\caption{Overall system architecture. (a) Server activated by AHLs'. (b) Green client activated. (c) Server re-activated by AHLs. (d) Red client activated.}
\label{fig:architecture}
\end{centering}
\end{figure}

In this Section we describe in detail the structure of our population-based client-server oscillator. We show how a three-strain community can, in principle achieve oscillatory behaviour (switching from red to green light output) in an autonomous, synchronous fashion. The basic architecture of our system, depicted in Figure ~\ref{fig:architecture}, is based on the ``client-server'' principle of modern computing, in which distributed {\it client} nodes communicate with a central {\it server}. In our system, we have one server strain and two client strains. We extend the analogy by considering the role of a {\it buffer}, which is the nutrient solution in which the cells live and grow (in computing, a buffer is a region of memory in which temporary data are stored). Signals are transmitted between client and server via this ``shared memory", through the actions of sensing and deposition. Each cell (``processor") also has its own private internal ``memory", corresponding to the space inside the membrane where local functions are performed. As each bacterium is, in effect, an independent processor, the success of our design relies on our ability to make all processors react {\it simultaneously} to external signals. 

Quorum-sensing (QS) \cite{atkinson2009quorum} has already been studied extensively in the context of synthetic biology \cite{Andrianantoandro:MolSystBiol:2006,Balagadde:MolSystBiol:2008,GarciaOjalvo:ProcNatlAcadSciUSA:2004}. This mechanism facilitates inter-bacterial communication via the generation and receiving of signal molecules \cite{fuqua1994quorum}. Most importantly, it enables a community-level response to emerge once a certain {\it concentration threshold} has been reached. It is this mechanism that we will use as the basis of the current study.
In what follows, there exist only four different signals, labelled AHLg, AHLr, AHLs and AHLs' (the precise definitions of each will be made clear later). Each cell/processor reacts not to the {\it absence} or {\it presence} of a specific AHL signal, but to the signal {\it level}, or concentration. For that purpose, a {\it threshold}, $\delta$, for input responses is defined in each cell. If the output, $O(B_i)$ produced by cell $B_i$ is required for activation of $B_j$, and some signal level is denoted by $|x|$, we assert that when $|O(B_i)| \geq \delta B_j$ then cell $B_j$ is activated. In this way, our model attempts to address one of the biggest problems inherent to single-cell circuits; stochastic expression noise due to inappropriate concentrations of signalling molecules.

In Figure ~\ref{fig:architecture}, we show the server and two clients; the server is {\it activated} by both AHLs and AHLs' (producing either AHLr or AHLg respectively); the {\it green} client is activated by AHLg, producing AHLs and green fluorescent protein, and the {\it red} client is activated by AHLr, producing AHLs' and red fluorescent protein. We can therefore see how this machine lies dormant until either AHLs or AHLs' is added to the nutrient, after which the system enters a period of oscillation (either red-green-red-... or green-red-green-... respectively). This is achieved by the server cells switching ``turns" between red and green client cells.  The intended behaviour of the server is shown in Table ~\ref{tab:server}; the important thing to note is that it acts as an XOR (exclusive OR) function, since it is only active when receiving a single input (i.e., when {\it one} of the clients is active). If, for some reason, either {\it both} or {\it none} of the clients are active, then the server should be {\it inactive} - this is central to the correct operation of the system.

We now describe in detail the internal structure of the client and server {\it E. coli} strains. In order to do so, we first specify the AHL molecules corresponding to the various signals within the system. These are listed in Table ~\ref{tab:signals}.

\begin{table}[b]
    \centering
    \begin{tabular}{ c  c  c  c  c}
    \hline 
    AHLs & AHLs' & AHLg & AHLr & S \\ \hline
    0 & 0 & 0 & 0 & 0 \\
    0 & 1 & 1 & 0 & 1 \\
    1 & 0 & 0 & 1 & 1 \\
    1 & 1 & 0 & 0 & 0 \\ \hline
    \end{tabular}
    \caption{Truth table for server strain.}
    \label{tab:server}
\end{table}

\begin{table}[]
    \centering
    \begin{tabular}{ c  c  c }
    \hline 
    Signal & System & Molecule \\ \hline
    AHLg & LasI/R & 3OC12AHL \\
    AHLr & Rh1I/R & C4AHL \\
    AHLs & LuxI/R & 3OC6AHL \\
    AHLs' & SinI/R & 3OC14AHL \\ \hline
    \end{tabular}
    \caption{Specific molecules corresponding to signals.}
    \label{tab:signals}
\end{table}

The four quorum sensing systems were chosen carefully, considering (1) sensitivity, (2) bacterial class, and (3) potential conflicts. In \cite{pai2009optimal} the four systems are grouped together in terms of their sensitivity, in \cite{lerat2004evolutionary} the systems are all characterised as being present in particular divisions of specific Proteobacteria, and in \cite{steindler2007detection} it is established that there exist no conflicts between the molecules sensed by the different systems. We therefore strongly believe that the systems we have chosen are appropriate.

\subsection{Server cells}

The server cells lie at the heart of the system, as they are responsible for implementing the core switching behaviour. In order to implement this, we use two {\it hybrid promoters}. These are promoters that are regulated by {\it two} inputs (one inducer and one repressor), and careful design allows them to be combined.  In order to activate the {\it red} output signal from the server, it first requires an input of AHLs.  

\begin{figure} 
\begin{centering}
\includegraphics[scale=0.6]{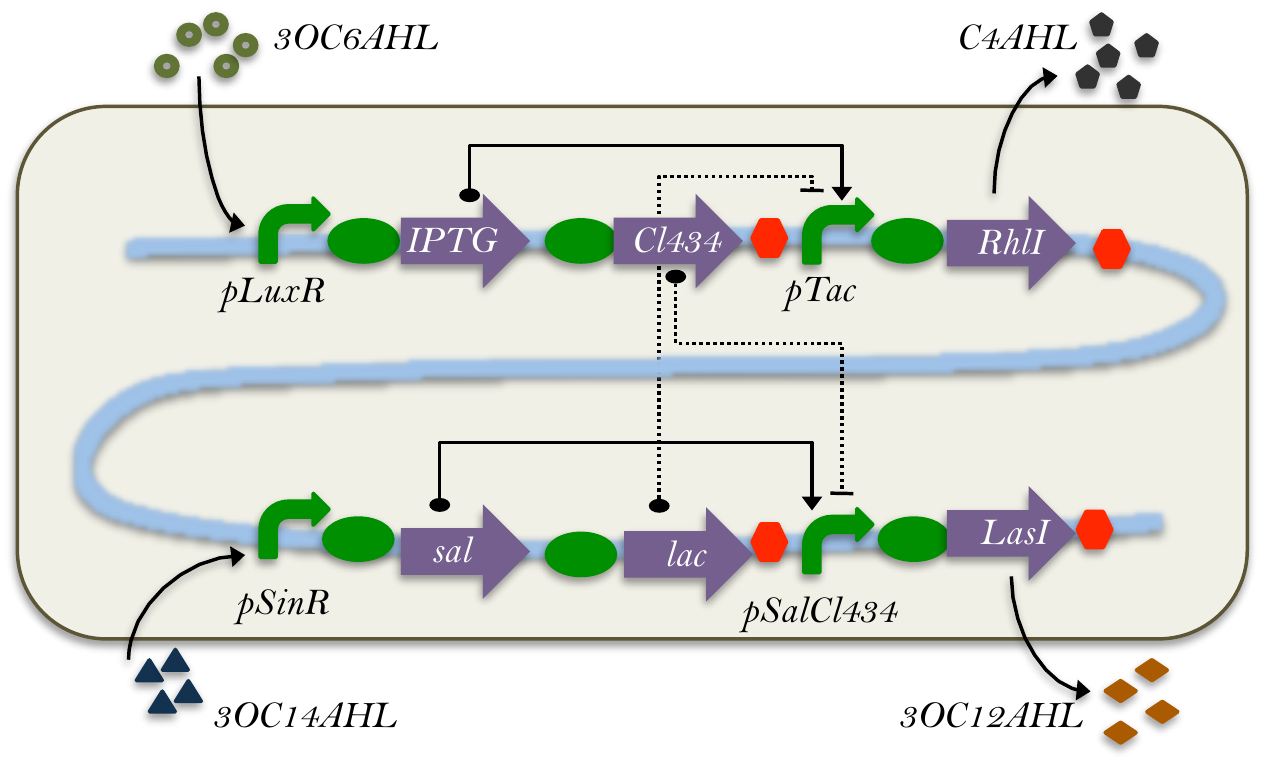}
\caption{Server cell internal architecture.}
\label{fig:server}
\end{centering}
\end{figure}

The detailed structure of the server is depicted in Figure ~\ref{fig:server}. When a server bacterium detects via its membrane that the concentration of AHL$_{s}$ molecules exceeds the input threshold for that $QS$ system, the inducible promoter \textit{pLuxRs} is activated. As a result of this, the two downstream structural genes are expressed. The production of IPTG molecules is used to stimulate a positive action in the {\it hybrid promoter} \textit{pTac} \cite{de1983tac} which, in turn, manages  the expression of AHL$_{r}$ molecules by using the gene \textit{luxIr}. At the same time, the expression product of the second gene, \textit{CI434}, represses the hybrid promoter \textit{pSalCI434}\footnote{Produced by the PKU Beijing team for IGEM2009, http://2009.igem.org}, so the production of AHL$_{g}$ is no longer possible due to the inhibitionof \textit{luxIg}. This general subsystem design is duplicated in the server to react symmetrically to each of the inputs the server may receive: AHL$_{s}$ and AHL$_{s'}$.

%http://2009.igem.org/Team:PKU_Beijing/Project/AND_Gate_2_Design

\subsection{Client cells}

\begin{figure} 
\begin{centering}
\includegraphics[scale=0.8]{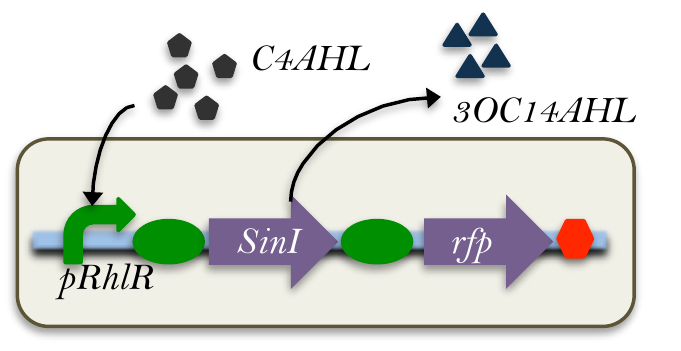}
\includegraphics[scale=0.8]{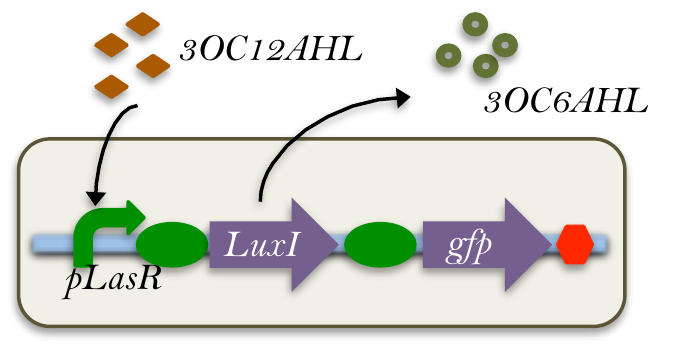}
\caption{Client cell internal architecture (Left: red, Right: green).}
\label{fig:clients}
\end{centering}
\end{figure}

The detailed structures of the client cells are shown in Figure ~\ref{fig:clients}. These cells have a much simpler design, because of the lack of synchronization requirements on clients within this model. In the case of the green client, when it senses a sufficient concentration of AHL$_{g}$ molecules in the environment to raise the threshold of the corresponding QS system, it activates the internal pathway that concludes with the expression of AHL$_{s}$ molecules and the reporter \textit(gpf). The first stage of this is the activation of the inducible promoter \textit{pLuxRg} which allows the transcription and translation of the genes \textit{LuxI} and \textit{gfp}. \textit{LuxIs} is used to produce end-turn signals (in this case, AHL$_{s}$), which are placed in the shared memory in order to notify the server that the green light that corresponds to half oscillation cycle has been satisfactorily expressed. The design of the red client is exactly the same, only with $SinI$ replacing $LuxI$, and $rfp$ being produced instead of $gfp$.

\section{Experimental results}

We now describe the results of simulation-based experiments to investigate the behaviour of both the individual components, and the client-server system as a whole. In order to achieve this, we run two sets of simulations, at different levels of detail. The first (micro-level) set investigates {\it intra-cellular} behaviour (i.e., at the internal level of gene regulation), and the second (macro-level) set of experiments assesses the effects of {\it inter-cellular} interactions. 

\subsection{Component testing: single-cell experiments}

\begin{figure} 
\begin{centering}
\includegraphics[scale=0.45]{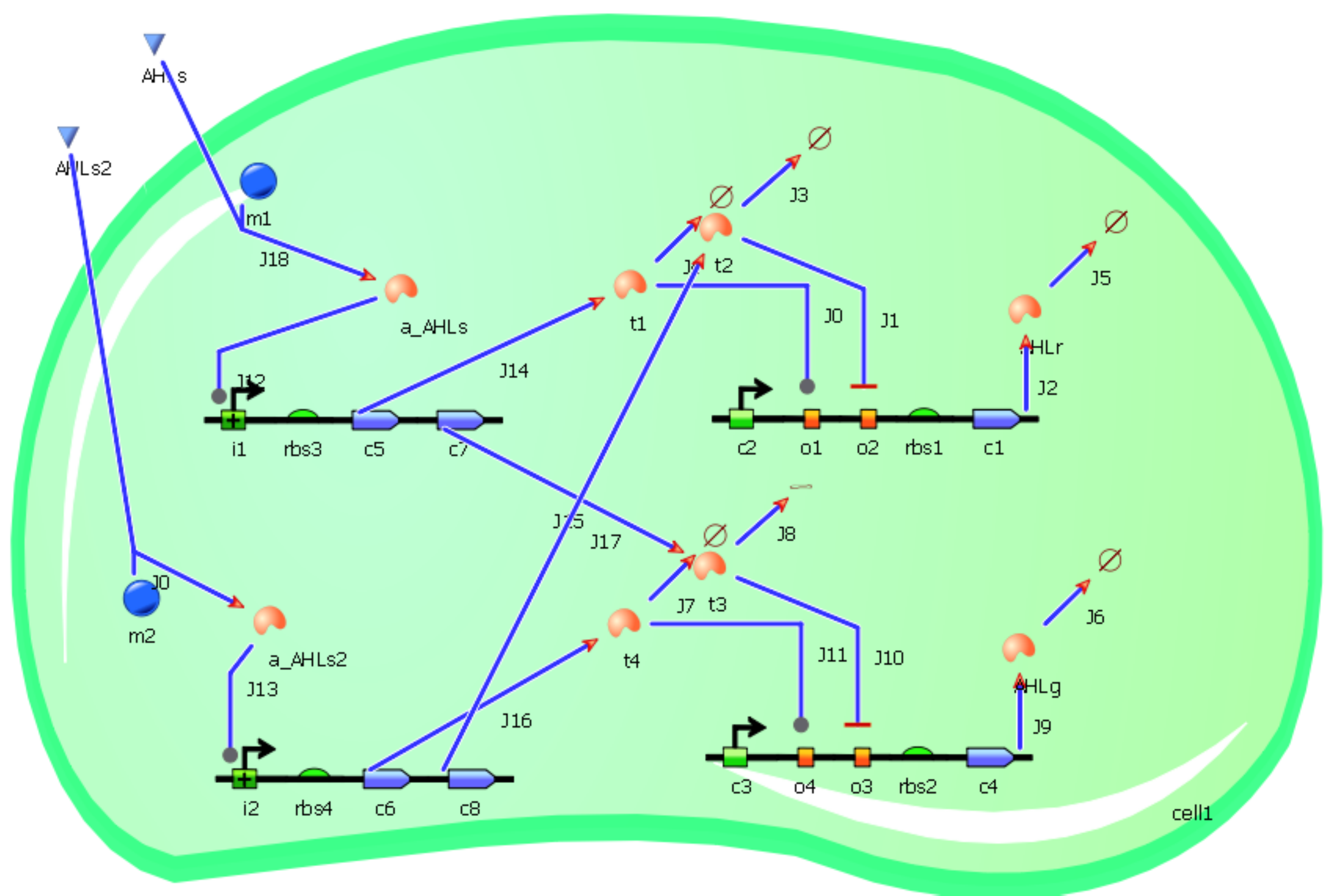}
\includegraphics[scale=0.45]{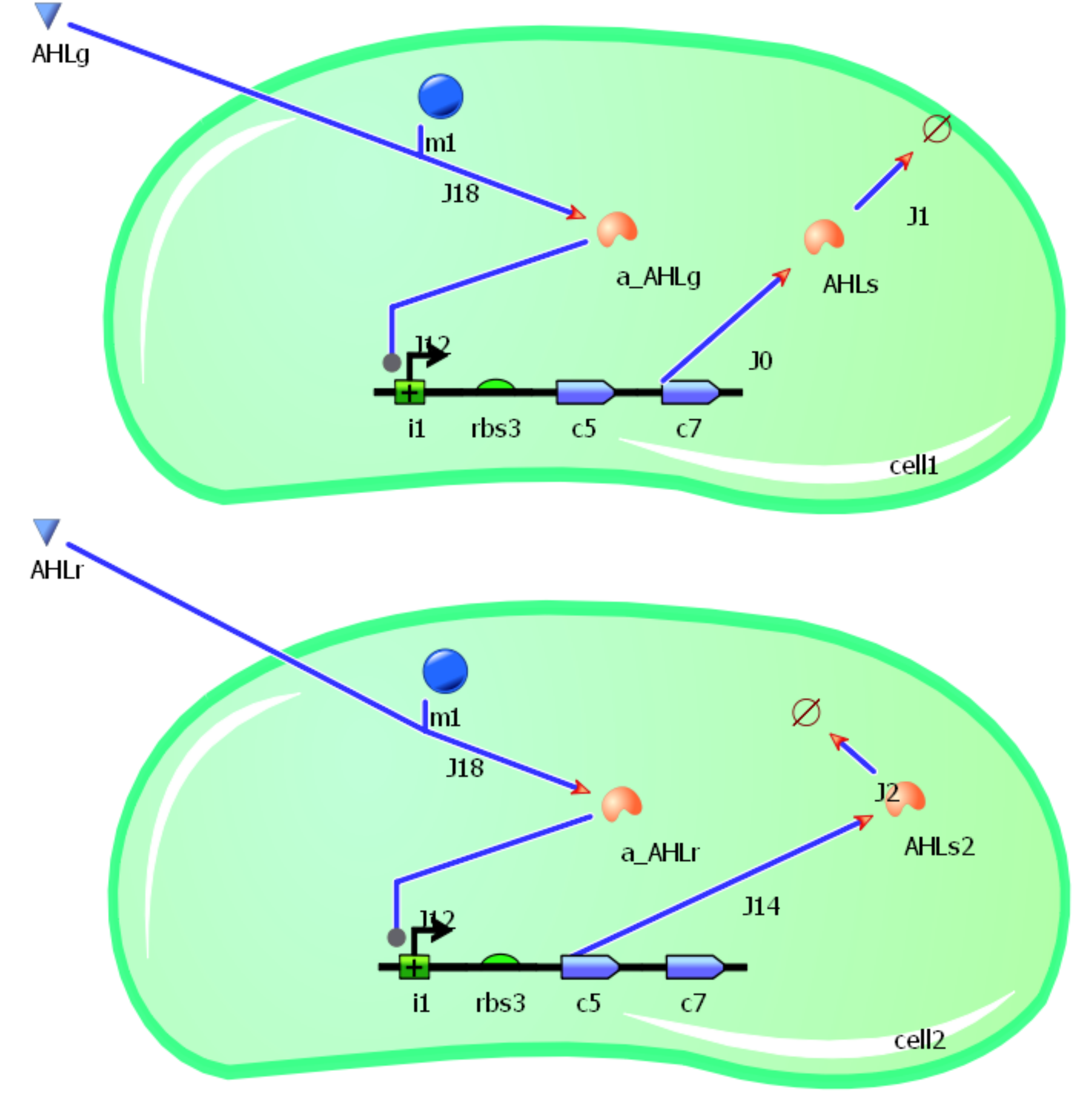}
\caption{Tinkercell schematics. Top: Server. Middle: Green client. Bottom: Red client.}
\label{fig:tinkercell}
\end{centering}
\end{figure}

For single component testing, simulations are performed using the Tinkercell \cite{Chandran:JBiolEng:2009} CAD tool to model the genetic networks of both the clients and the server. This simulation environment provides a framework for the study of the dynamical behaviour of genetic circuits. The user specifies the different components of the circuit (either taken from a standard library or defined on an {\it ad hoc basis}) and their connectivity, as well as external inputs and outputs. The key detail lies in the correct specification of the nature of connections between components. These connections are represented by ODEs, which draw their terms from the individual components. Tinkercell then solves the set of equations on behalf of the user, producing dynamical time-series plots. The Tinkercell schematics for the server and clients are depicted in Figure ~\ref{fig:tinkercell}.  The server hybrid promoters (c2 and c3 in figure \ref{fig:tinkercell}) have two different operator regions where the molecules can bind, one for inducible forces and one for repressible forces. When AHLs exceeds the membrane threshold it binds molecules $m1$ (which are always present in the cell) , and creates $a\_AHLs$ (active AHLs molecules). Those molecules induce promoter $i1$ activity. The same principle applies for AHLs2 (i.e., AHLs', the labelling is simply an restriction of the software).

Tinkercell provides default equations by which various standard components are connected, and no changes were made to any of these. All concentrations are measured in $\mu M$ units. However, it is relevant to remark that the strength of the RBSs and the promoters has been increased to a very high value in order to detect possible irregularities in the system (RBS\_strength = 20; Promoter\_strength = 4). That is why in this single component specification, we observe high values for output responses. In this way, we artificially amplify both ``good" and '``bad" signals, in order to easily detect aberrant behaviour. However, this does not {\it change} in any way the overall behaviour of the system. All the intermediate transcription factors of the system are set up to 0 $\mu M$ at the beginning of the simulation, and we include a degradation reaction for each.

% (important not to assign that rate to the specific R proteins).

\begin{figure}[]
   \centering

       \includegraphics[scale=0.6]{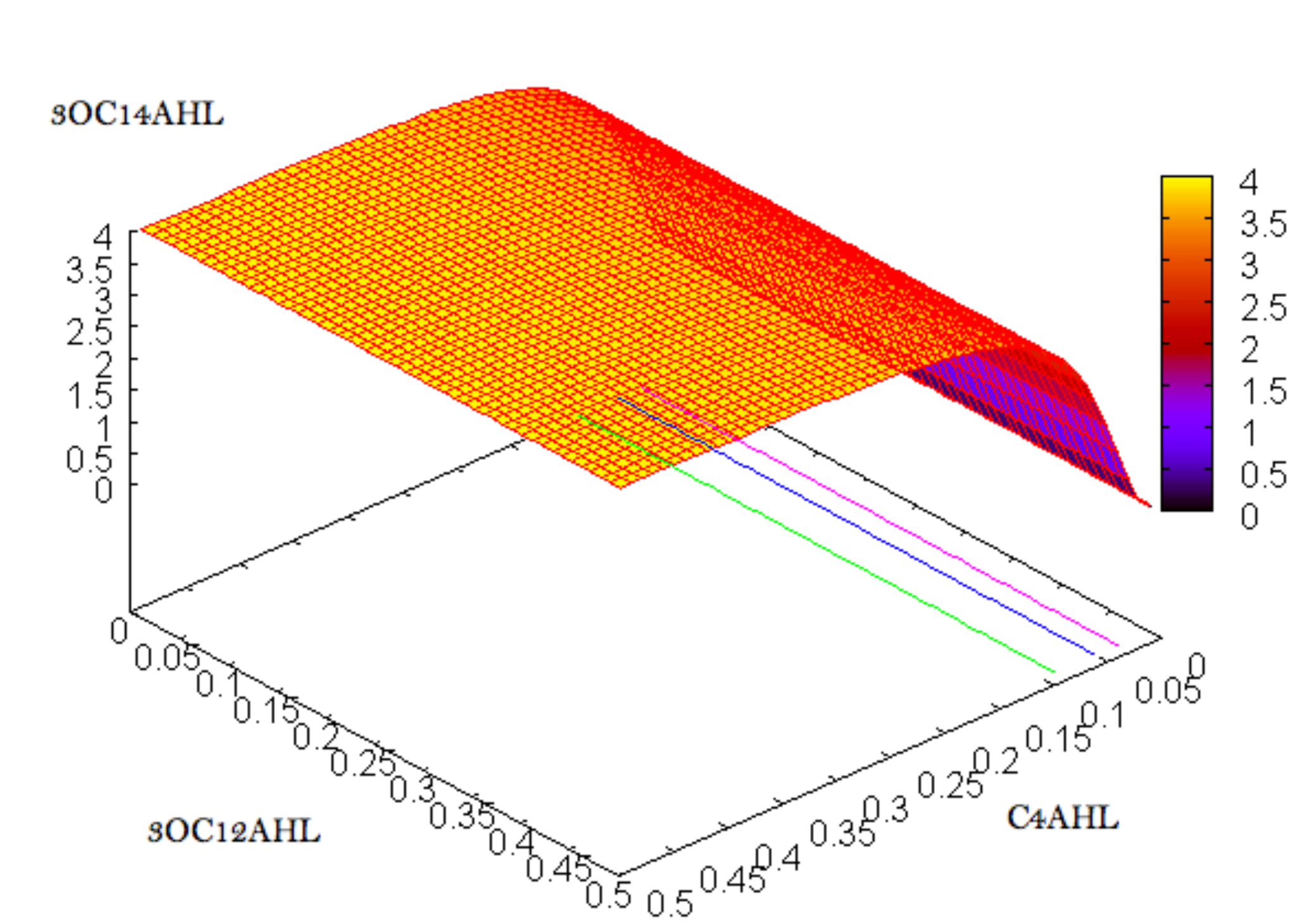}      
        \includegraphics[scale=0.6]{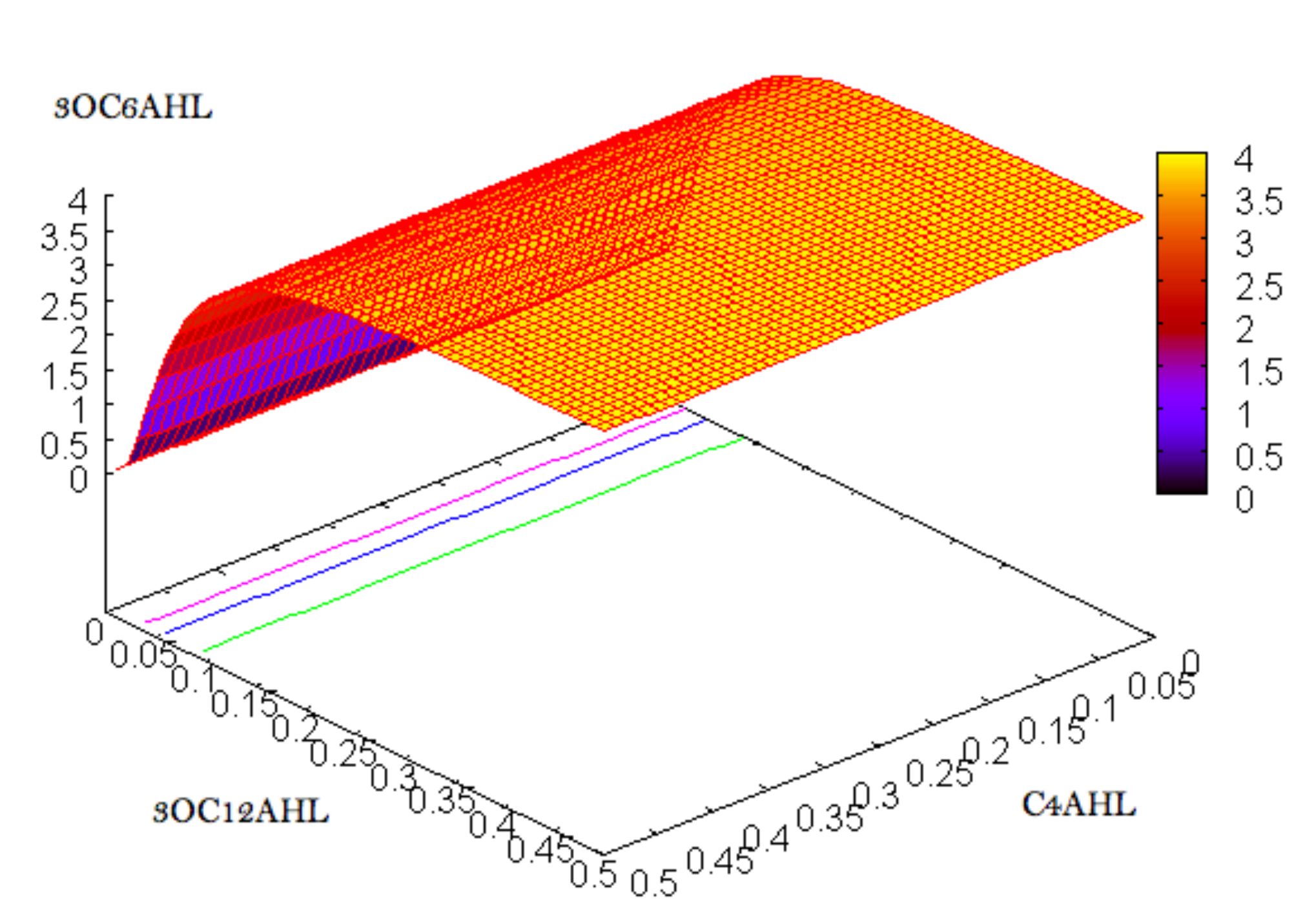}
   \caption{Client behaviour. Top: Red client. Bottom: Green client.}
   \label{fig:sim10}            
\end{figure}

We now show the results of single-cell simulations for clients and server. We vary the inputs to each in order to confirm that they only produce a threshold output at the correct concentration levels. Figure \ref{fig:sim10} shows the behaviour of the two client strains for different concentrations of input molecules.  Molecule C4AHL, which is the red-specific turn signal, induces the expression of ClientR output. It is very important to note that the concentration of those molecules which raise thresholds is denoted by a value of zero for C4AHL. That is the moment in which ClientR's turn begins, by launching the cascade of its quorum sensing system to produce 3OC14AHL (end turn signals) and red fluorescent proteins (for system output). When C4AHL molecules exceed a concrete saturation level (higher than the threshold), the production of ClientR's output reaches its maximum and the client reaches an steady state. We note that ClientR's operative machinery is completely indifferent to green-specific turn signals (3OC12AHL). A similar behaviour affects the green client strain with opposite molecular perception.  

We now describe the simulation tests of the server component of the model. Since noise-attenuation inside the server is one of the most important features of the model, this component is tested extensively with a wide range of values for the degradation rates of the four transcription factors connected to the hybrid promoters. The overall objective is to ensure that the server only ever emits a single signal during any one '``turn". We first performed a set of runs using random values for each degradation rate, which confirmed that the system performed correctly and robustly. We were only able to obtain incorrect behaviour for the system by using abnormal and unrealistic degradation rates. In what follows, we use the default degradation rate (-1) for each protein.

\begin{figure}[]
   \centering
        \includegraphics[scale=0.6]{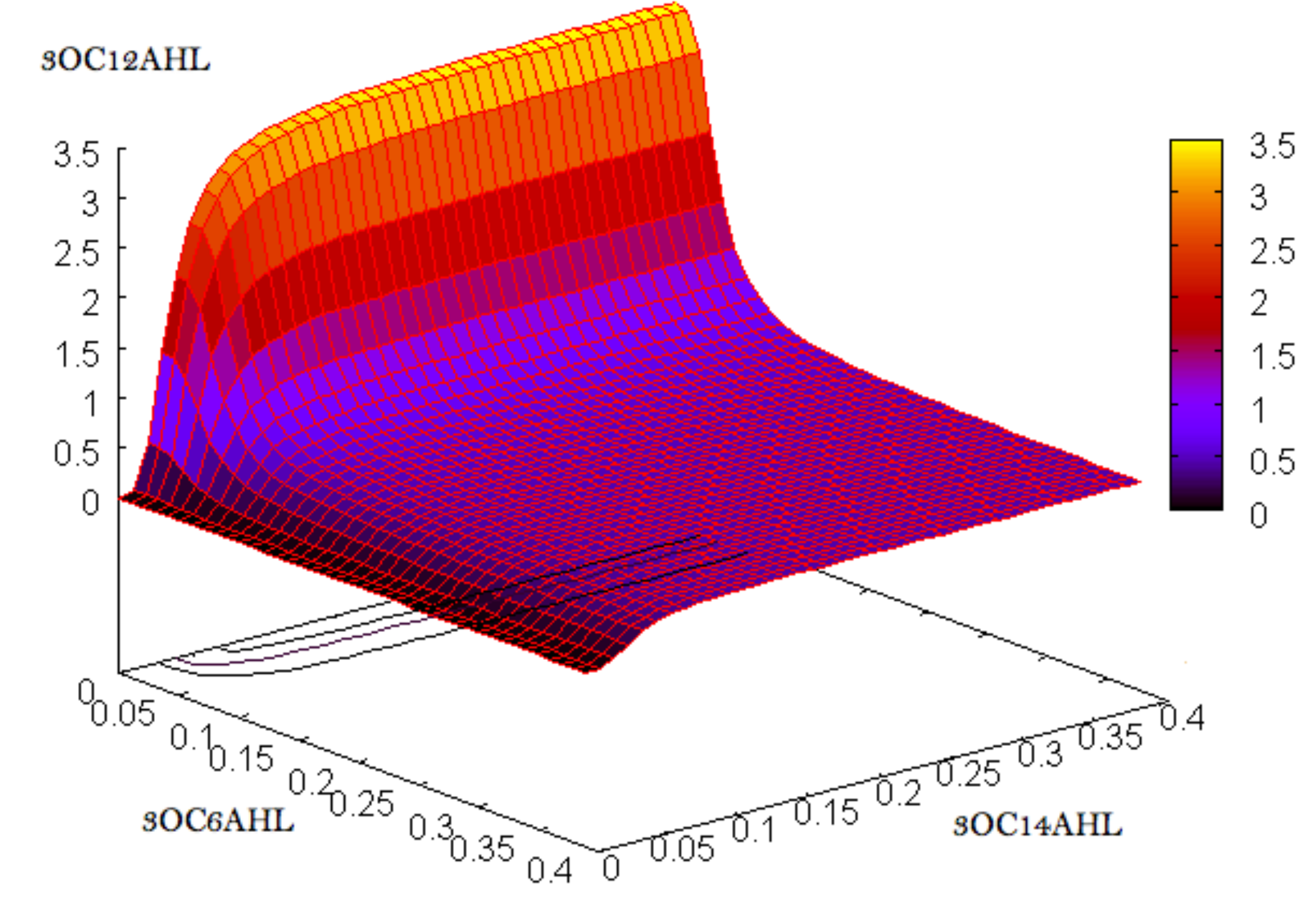} \\
        \includegraphics[scale=0.6]{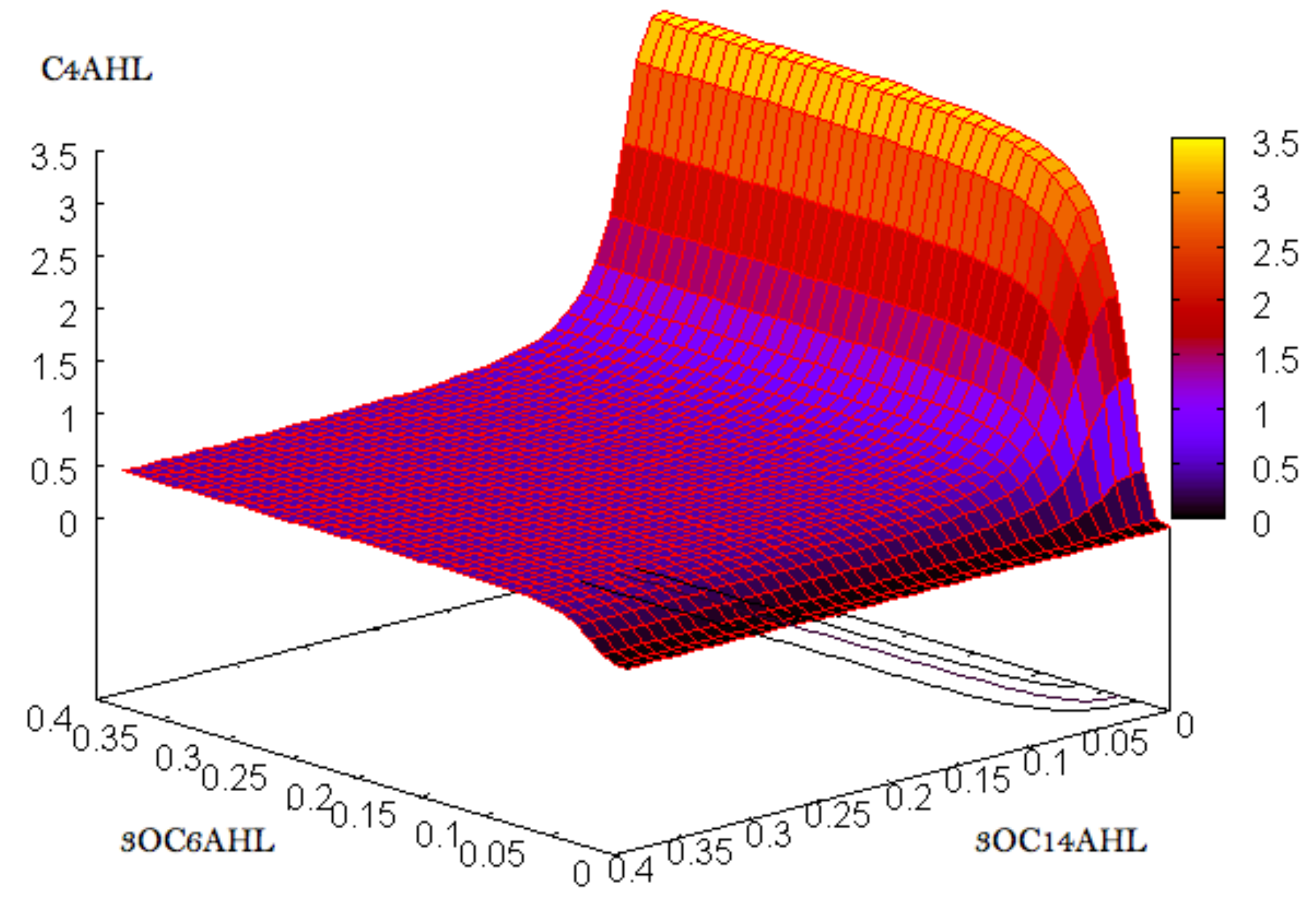}
   \caption{Server behaviour. Top: Green turn output. Bottom: Red turn output.}
   \label{fig:sim1}            
\end{figure}

Again, we vary the concentrations of both inputs in order to confirm that the server yields the correct output.
The graphs in Figure \ref{fig:sim1} illustrate the operation of the server during each ``turn". Figure \ref{fig:sim1} (top) depicts the situation in which the server receives red-specific {\it end turn} signals (3OC14AHL) and produces green-specific turn signals (3OC12AHL). In order to make this graph clear, two important features must be highlighted. Firstly, a zero value for an input signal (x and y axes) represents, as in the client graphs, the instant at which the concentration of the corresponding molecule in the nutrient solution is sufficient to exceed the membrane threshold so that the server can start. However, the most relevant attribute of the server is the interaction of red and green end turn signals inside the cell. As we observe in graph \ref{fig:sim1} (top), red-specific end turn signals activate the expression of the green turn {\it if and only if} green-specific end turn signals (3OC6AHL) are {\it not present} inside the server (or at a very low concentration). This characteristic aids noise attenuation and, therefore, the correct operation of the model in case of saturation of the system, by causing a halt state.

These simulation tests confirm, in principle, the correct functioning of the individual system components. We now describe the results of full-system simulations to assess the overall behaviour of our client-server model. 

\subsection{System-level experiments}

Having confirmed the correctness of the individual component designs, we now seek to investigate the overall system behaviour. In order to achieve this, we treat each of the three components as a ``black box" within a simulation based on Michaelis-Menten kinetics \cite{cornish1995fundamentals}  and performed using the BioBrick library for Simulink (cite) . The fundamental schematics for each component are the same as in the Tinkercell simulation; the only difference lies in the manner in which the system equations are solved by the software. In both of them, abiological processes are represented by a set of ODEs:

\begin{multicols}{2}
\begin{equation}
\mu(t) = p(t) - \delta\mu(t) \label{trasc}
\end{equation}

\begin{equation}
p(t) = \beta {FT \over \gamma + FT} + \alpha \label{activ}
\end{equation}

\begin{equation}
\eta(t) = \lambda\mu(t) - \delta_{p}\eta(t) \label{transl}
\end{equation}

\begin{equation}
p(t) = \beta {1 \over \gamma + FT} + \alpha \label{rep}
\end{equation}
\end{multicols}

Equation \ref{trasc} describes the genetic transcription process $\mu(t)$, where $p(t)$ is the production rate and $\delta$ the linear degradation rate. Equation \ref{activ} represents the action of an inducible promoter (needed for the QS system), where $FT$ is the transcription factor, $\gamma$ a saturation constant, $\beta$ the maximum transcription rate and $\alpha$ a default transcription constant. For concluding protein production, the model obeys equation \ref{transl} where $\lambda$ is a constant and $\delta_{p}$ the degradation rate of the final output $\eta(t)$. Finally, equation \ref{rep} denotes a repressible promoter, which is used to build in Simulink the negative operator of the hybrid promoters.

While Tinkercell uses an integrator (CVODE) to build the final system, the BioBrick library eliminates nonlinearities by linearising the ODEs using a frequency (Laplacian) domain.

%the only difference lies in the manner in which the circuits are simulated. In Tinkercell, the system is translated into a set of ODEs, and in Simulink (which is a library for MATLAB), the system is simulated numerically.

We perform three sets of experiments to assess the system-level behaviour. In the first, we assume an idealised situation, with zero cell growth. The second set of experiments investigates a rather more realistic situation, which accounts for cell growth. The third set of experiments simulates the situation in which one client has a different signal degradation rate to the other (in order to both confirm correct system behaviour, and to open up the possibility of different oscillatory patterns). 

Before describing the experiments, we first make explicit some assumptions. Given that the minimal concentration ($c_{m}$) of molecules that a strain can produce is equal to 0, and the maximal concentration ($c_{M}$) is equal to a predefined value (5 in the selected simulations), concentration values within the simulation lie in the interval [0...5]. (percentage proportion). Within that interval, the numeric value of the membrane threshold can be specified within the interval [$\simeq$ 0.7 ... $\simeq$ 3.7] which represents 60\% of the global molecule production (or 2\% if repression forces are not considered). In this way, we hope to obtain robust system behaviour across a range of environmental conditions. In what follows, time units are dimensionless, as this parameter is dependent on the concentration value. We note that, in all of the graphs, the level of green fluorescence is coupled to the concentration of AHLs, and the level of red fluorescence is coupled to the concentration of AHLs'.  We now investigate three different scenarios: (1) an idealised situation, with a static population, (2)
a more realistic situation, with a {\it growing} population, and (3) the situation where clients have different degradation rates associated with them.

\subsubsection{Scenario 1: Static population}

\begin{figure}[h!]
   \centering       
        \includegraphics[scale=0.6]{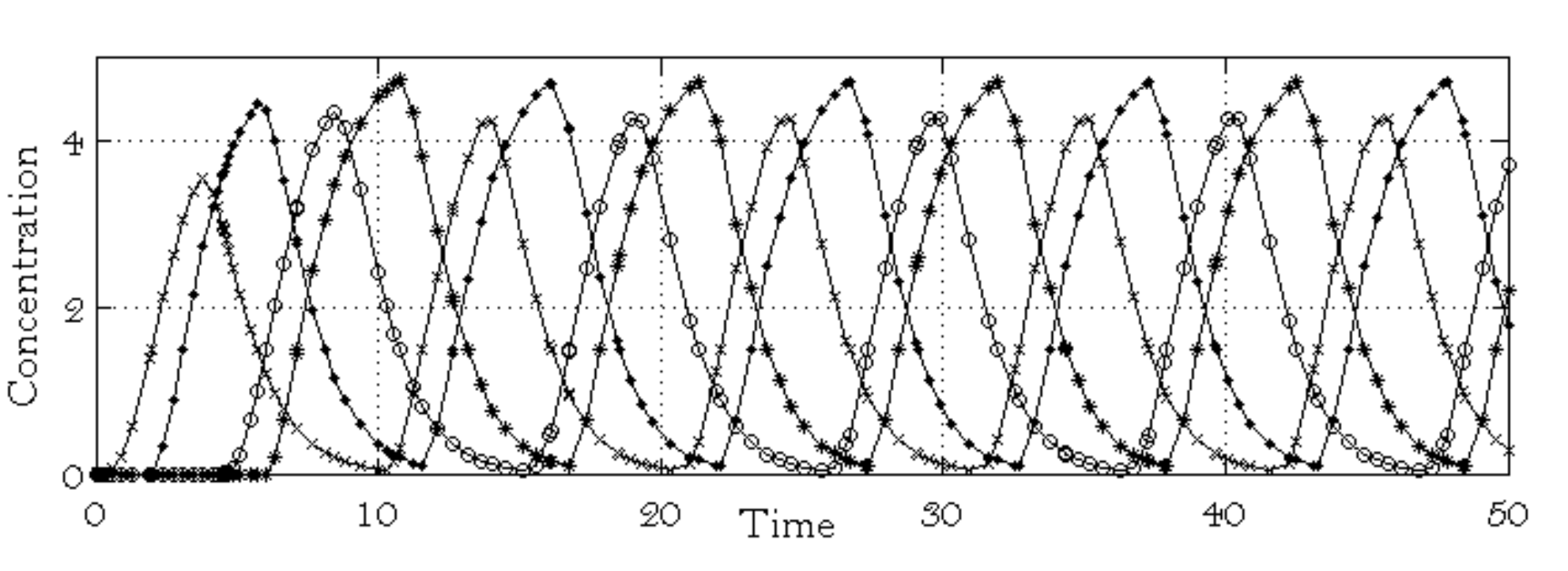}
   \caption{Simulation results, static population. Key: AHLr($\times$), AHLg($\circ$), AHLs'($\bullet$), AHLs($\ast$).}
   \label{fig:stathigh}            
\end{figure}

We first study the simple situation in which the bacterial population is static (i.e., there is no growth). This is an idealised example, in which we effectively simulate a system containing a single server, and one of each client. The results are depicted in Figure ~\ref{fig:stathigh}.

We observe the expected pattern of oscillation, starting with the red client. This turn is denoted by a rise in concentration of AHLr. The important thing to note is that the concentration of AHLg is basal until the end of the red turn is signalled by a rise in concentration of AHLs'. This triggers the green client's turn, signalled by a rise in concentration of AHLg, the end of which is indicated by a rise in concentration of AHLs. This last rise then triggers the activity of the red client once more, and the cycle continues.
 
\subsubsection{Scenario 2: Growing population}

The second experiment considers the situation in the different strain populations observe the usual growth patterns. Rather than explicitly simulating each bacterium or coarse-grained sub-population (which would require a spatially-explicit model, which is beyond the scope of the current study), we reproduce the effect of a growing strain by multiplying the output of a single population member according to the standard log phase growth function.

\begin{figure}[h!]
   \centering       
        \includegraphics[scale=0.6]{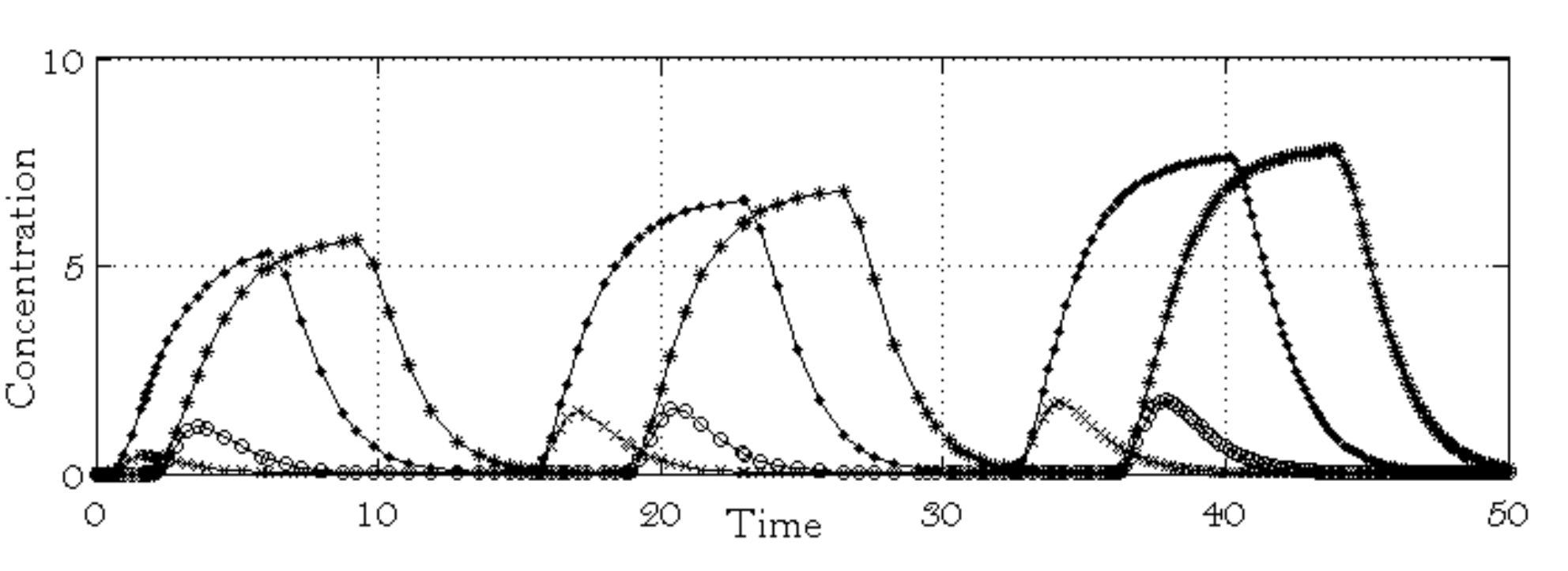}
   \caption{Simulation results, growing population. Key: AHLr($\times$), AHLg($\circ$), AHLs'($\bullet$), AHLs($\ast$). }
   \label{fig:dyn}            
\end{figure}

Again, we observe the expected system behaviour, but see a gradual rise in the overall maximum concentrations of AHLs and AHLs', which is consistent with population growth. The concentration levels of AHLg and AHLr are lower than in the previous experiment because they correspond to ``start" signals to the clients. The server therefore has the capability of ''pausing" the oscillation until the correct conditions are in place.

\subsubsection{Scenario 3: Differential client behaviour}

\begin{figure}[h!]
   \centering       
        \includegraphics[scale=0.6]{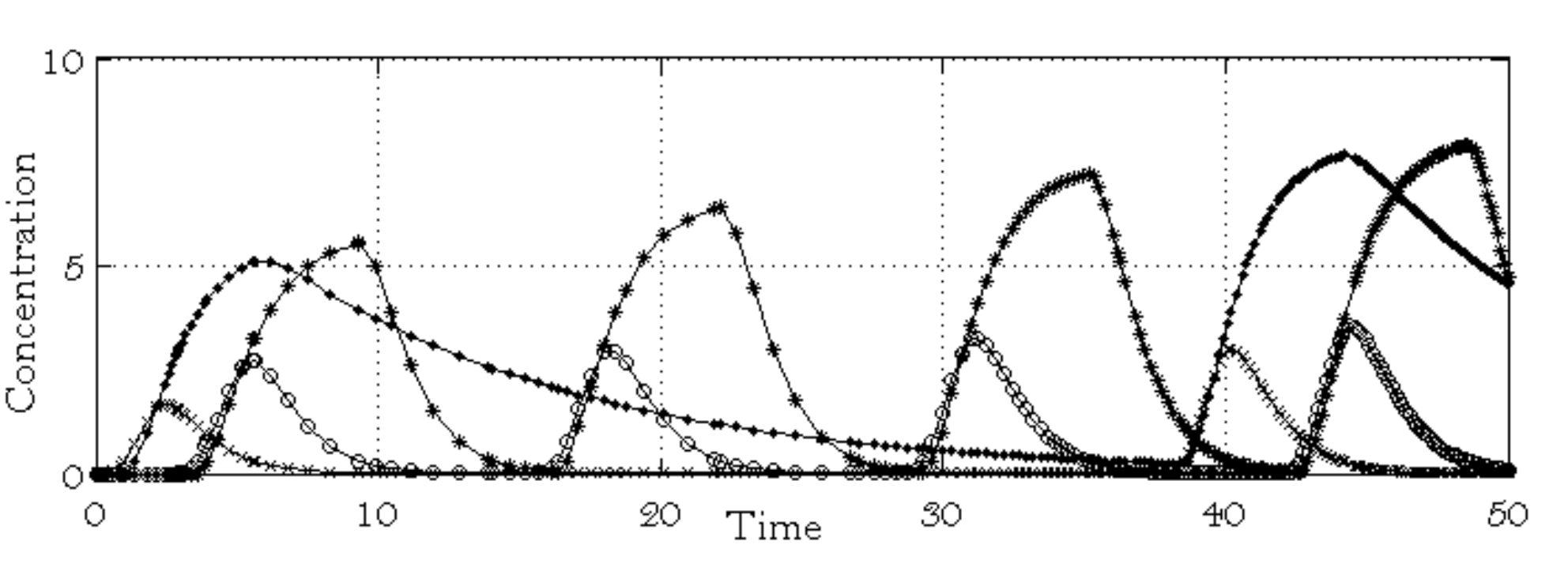}
   \caption{Simulation results, differential client behaviour. Key: AHLr($\times$), AHLg($\circ$), AHLs'($\bullet$), AHLs($\ast$).}
   \label{fig:dyndeg}            
\end{figure}

In the final experiment we investigate the effect of differential behaviour of the client strains. We modify the degradation rate of the ``red off" signal AHLs', so that it is removed much more slowly from the system. In Figure ~\ref{fig:dyndeg} we observe the effect of this change. The red client finishes its turn, but the end-turn molecules degrade more slowly, and are therefore still present in the shared memory. The server notices this red end-turn, and repeatedly yields the turn to the green client until the AHLs' molecules disappear. The system can therefore adapt its behaviour to this new situation. We do not observe a green-red-green-red pattern, but instead see red-green-green-green-green-red.... Importantly, the system dynamically reconfigures the oscillation pattern in a manner that is completely consistent with correct architectural behaviour.

\section{Conclusions}

In this paper we presented a design for a population-based cellular oscillator, which uses quorum sensing-based signalling within a client-server model. Simulation studies of our design suggest that it is realistic and robust to fluctuations in environmental conditions. Such systems will become increasingly important for synthetic biology, as the field seeks applications in (for example) distributed bio-sensing or tissue engineering. Future work will focus on refinements of the model, as well as its experimental validation.

\bibliography{oscillator}

\end{document}